\documentclass[
  apj,
  appendixfloats,
  iop,
  numberedappendix,
  tighten,
  twocolappendix,
]{emulateapj}

\usepackage{amsmath}
\usepackage{amssymb}
\usepackage{microtype}
\usepackage[
  allcolors=blue,
  colorlinks,
  pdftex,
  pdfauthor={Manu Mannattil, Himanshu Gupta, and Sagar Chakraborty},
  pdftitle={Revisiting Evidence of Chaos in X-ray Light Curves: The Case of GRS 1915+105},
  pdfsubject={High energy astrophysical phenomena; chaotic dynamics},
  pdfkeywords={chaos; methods: data analysis; X-rays: binaries; X-rays: individual (GRS 1915+105)}
]{hyperref}

\usepackage{txfonts}
\usepackage{bm}
\renewcommand{\mathbf}{\bm}

\newcommand\blfootnote[1]{%
  \begingroup
  \renewcommand\thefootnote{}\footnote{#1}%
  \addtocounter{footnote}{-1}%
  \endgroup
}

\submitted{}

\shorttitle{Revisiting evidence of chaos in GRS~1915+105}
\shortauthors{Mannattil, Gupta, \& Chakraborty}

\begin{document}

\title{Revisiting evidence of chaos in X-ray light curves: the case of GRS~1915+105$^\star$}
\author{
  Manu Mannattil, Himanshu Gupta, and Sagar Chakraborty
}
\affil{
  Department of Physics, Indian Institute of Technology Kanpur, Uttar Pradesh 208016, India\\
  \href{mailto:mmanu@iitk.ac.in}{mmanu@iitk.ac.in},
  \href{mailto:hiugugpta@iitk.ac.in}{hiugupta@iitk.ac.in},
  \href{mailto:sagarc@iitk.ac.in}{sagarc@iitk.ac.in}
}

\begin{abstract}
  Nonlinear time series analysis has been widely used to search for signatures of low-dimensional chaos in light curves emanating from astrophysical bodies.
  A particularly popular example is the microquasar GRS~1915+105, whose irregular but systematic X-ray variability has been well studied using data acquired by the \emph{Rossi X-ray Timing Explorer}.
  With a view to building simpler models of X-ray variability, attempts have been made to classify the light curves of GRS~1915+105 as chaotic or stochastic.
  Contrary to some of the earlier suggestions, after careful analysis, we find no evidence for chaos or determinism in any of the GRS~1915+105 classes.
  The dearth of long and stationary data sets representing all the different variability classes of GRS~1915+105 makes it a poor candidate for analysis using nonlinear time series techniques.
  We conclude that either very exhaustive data analysis with sufficiently long and stationary light curves should be performed, keeping all the pitfalls of nonlinear time series analysis in mind, or alternative schemes of classifying the light curves should be adopted.
  The generic limitations of the techniques that we point out in the context of GRS~1915+105 affect all similar investigations of light curves from other astrophysical sources.
\end{abstract}

\blfootnote{$^\star$\,Compared to the published version, this version contains additional discussion and review of previous studies.}


\section{Introduction}

GRS~1915+105 \citep{castro-tirado92,belloni00,fender04}, like Cyg~X-1 \citep{lochner89,unno90,timmer00}, is a binary star system where one member is a black hole onto which the matter of the partner accretes.
Similar binary systems are Her~X-1 \citep{voges87,norris89} and Sco~X-1 \citep{mendez00,karak10}, where, however, the compact accreting star is a neutron star.
These binary systems depict variability in their emitted X-ray photon counts over time scales ranging from milliseconds to months.
Although details are not often known, it is customary to attribute this variability to the fluctuations in the accretion rates, hydrodynamic instabilities in the inner regions of the accretion disks, etc.
Hence, the X-ray light curves emitted by such systems carry information about the geometry of the accretion disk and the physical processes involved in the emission process.
X-ray variability is also inherent to emissions by accreting active galactic nuclei (AGNs), e.g., in Ark~564 \citep{gliozzi02}, and 3C~390.3 \citep{gliozzi06}.
Among AGNs, blazars like W2R~1926+42 \citep{bachev15}, whose emissions are arguably emanating from relativistic jets, show variability in the optical range.
Then there are variable stars, e.g., R~Scuti, R~UMi, RS~Cyg, V~CVn, UX~Dra, SX~Her, etc.~\citep{buchler96,buchler04}.
The fluctuating and apparently random light curves from all these and similar astrophysical bodies carry vital information about the sources' emission mechanisms.
Moreover, they can be used to classify astrophysical systems into equivalent classes.
A recent example is that of strange nonchaotic stars: Kepler spacecraft observations confirm that several stars pulsate with multiple frequencies.
Some like the RRc~Lyrae star KIC~5520878 \citep{lindner15}, pulsate with two principal frequencies that are nearly in the golden ratio.
Apparently, the light curve of this star shows strange nonchaotic behavior, making it a prototype of a new class.
Even a single source can be categorized into different temporal variability classes in a model-independent way.
For example, \cite{belloni00} showed that GRS~1915+105 has 12 well-known classes (also cf.~\citealt{hannikainen05} and \citealt{choudhury07}).
Though GRS~1915+105 displays several variability modes, this particular classification allows one to account for all the variations in terms of transitions between only three main spectral states.
Although \cite{neilsen11} showed that the mass-loss rate in the accretion disk wind may be adequate to create long-term oscillations in the accretion rate causing state transitions, a fully detailed explanation for GRS~1915+105's energy-dependent variability is still missing.

There are a plethora of systems similar to the select few that have been explicitly named in the preceding paragraphs.
Actually, we have consciously highlighted those specific systems that have been investigated using the nonlinear time series analysis technique that is the main subject of this paper.
An astrophysicist analyzing a light curve wants to understand the mathematical nature of the simplest possible model that can generate such a light curve.
Of course, understanding any physical phenomenon in its entirety is a difficult and an impractical endeavor.
The practiced approach toward solving many problems in physics, therefore, is to understand them through a mathematical model.
The model is usually invented after stripping down the physics to its barest minimum, so that the main reason behind the phenomenon is revealed in the simplest possible manner.
Consequently, simpler low-dimensional models are of paramount practical interest to physicists, astrophysicists, and researchers in other disciplines.
The following discussion elaborates more about this in the context of the astrophysical scenario being discussed in this paper.

The energy supply of an accreting astrophysical body comes from the release of gravitational energy due to matter inflow onto its surface.
The fact that a substantial fraction of the emission is in the X-ray and gamma ray bands hints that the emission is happening close to the surface.
We thus expect the X-ray variability to provide clues to activities occurring in the vicinity of the accreting body.
These X-ray emissions have aperiodic variability and generally featureless self-similar power spectra.
This could be due to short-living uncorrelated elements, e.g., random reconnections appearing in the large-scale magnetic field lines, or randomly generated shocks in the outflowing (or inflowing) gases.
In such a scenario it is customary to use stochastic processes, such as shot noise processes, to model the phenomenon.
Intriguingly, this could also be due to the coherent variability of a cloud of electrons, which may be modeled using a geometrically thick, optically thin disk type model \citep{czerny97}.
In such situations, the variability can be characterized using a dynamical system capable of exhibiting deterministic chaos.
For example, the AGN in NGC~4051 has been reported to emit chaotic light curves \citep{lehto93}.
Naturally, detection of chaos in the light curves can settle the issue of which model---chaotic or stochastic---is more apt.
Furthermore, \cite{winters03} studied flow turbulence due to magnetorotational instabilities occurring in the accretion disk and calculated quantitative chaos parameters, like the largest Lyapunov exponent.
Thus, if the X-ray variability is indeed coupled with the instabilities in the accretion disk, then the possibility of determinism (low-dimensional chaos) in the light curves opens up.
Hence, the analysis of the light curves using nonlinear techniques helps us to put restrictions on theoretical models in a model-independent way.
Additionally, it can also put some constraints on the free parameters of the models.
Owing to such an analysis, one also gains more (although indirect) information about the dynamical structure of the underlying system producing the signal.
Hence, it can inspire new models or at least provide additional tests for existing ones.

The concept of chaos, as used above, is straightforward.
Chaos is found in deterministic nonlinear dynamical systems whose phase space trajectories are ultrasensitive to initial conditions, i.e., infinitesimally nearby initial conditions evolve in drastically different ways in the long run.
This leads to the impression that a chaotic system is behaving `randomly' even when the system is completely deterministic: given a unique past state, there is always a unique future state.
Thus, technically, the word `random' should be used to describe nondeterministic stochastic processes and not chaotic ones.
It must also be emphasized that though nonlinearity is a necessary condition for a system to be chaotic, it is not a sufficient one.

Deterministic dynamical systems, including chaotic ones, are often best characterized by defining an attractor.
While researchers have debated the usefulness of the various definitions of an attractor, for our purpose, the following operational definition \citep{eckmann85} suffices: given a $d$-dimensional autonomous system, its attractor is a set of points in the phase space such that \emph{almost} every initial condition near this set reaches it asymptotically.
Arguably, the most exotic attractor is the chaotic attractor, which is mostly strange, i.e., it has nonintegral fractal dimensions.
By definition, existence of a bounded attractor, even if chaotic, means that the dynamics is confined and in a sense stable (i.e., not explosive) in the long run.

Two very different characterizations of a chaotic attractor are possible: we call them the dynamic characterization and the geometric characterization.
As a chaotic orbit is wandering in an attractor, on average it tends to have exponential local divergence from other nearby trajectories.
This divergence is quantified by the maximum Lyapunov exponent ($\lambda_{\max}$), whose positivity is taken as the most convenient definition for chaos \citep{faust94}.
Thus, the specification of the maximum Lyapunov exponent (and other Lyapunov exponents\footnote{In general, in a $d$-dimensional autonomous dynamical system there are $d$ Lyapunov exponents that measure how on an average a hypersphere of initial conditions evolves into a hyperellipsoid through exponential contractions or expansions along the principal axes.}) for an attractor is part of its dynamic characterization.
In contrast, geometrical characterization is typically done by finding the fractal dimension of the attractor, which could either be the box dimension ($D_0$), the information dimension ($D_1$), or the correlation dimension ($D_2$).
In full generality, one can also define generalized dimensions ($D_q$; $-\infty<q<\infty$) to quantify the static geometry of a chaotic attractor \citep{feder88}.
Both characterizations are related in some cases via the Kaplan--Yorke conjecture \citep{Kaplan1979}.
Also, the loss of information due to the folding of trajectories in the phase space is quantified by the Kolmogorov--Sinai entropy $K$, which is closely related to the Lyapunov exponents, for example, by Pesin's entropy formula \citep{pesin97}.

Now consider a typical light curve under discussion. The light curve, obtained through a photon counting process, is always noisy.
Even if it were possible to filter out all other types of noises, Poisson noise due to the quantum nature of light can never be fully removed.
Fluctuating photon counts that detectors aboard satellites such as the \emph{Rossi X-ray Timing Explorer} (\emph{RXTE}) register over time can be interpreted as (within some proportionality constant) the intensity $I(t)$ of light corrupted with noise.
In principle, under the mask of the noise, one hypothesizes a \emph{classical dynamical system} with $I(t)$ as one of the many possible functions of some phase variables.
The main question that nonlinear time series analysis attempts to answer is whether this hypothesis is true.
In the jargon of probability theory, establishing determinism thus necessarily amounts to proving that the nonhomogeneous Poisson counting process with rate $I(t)$ generating the light curve is not a Cox process---also known as a doubly stochastic Poisson process \citep{cox80}.
To summarize, we want to find whether $I(t)$ can be generated by a deterministic dynamical system asymptotically, and if so, then what are the values of the invariants (e.g., $\lambda_{\max}$, $D_2$, $K$, etc.; also see the next section) that characterize its attractor.

Thus, like many other researchers, the main question we ask in this paper is, can some light curves collected from GRS~1915+105 be modeled as a solution of a deterministic dynamical system?
An answer to this question is traditionally sought by employing the methods of nonlinear time series analysis based on the theory of dynamical systems.
The first investigation of this kind was carried out by \cite{misra04,misra06}, who proposed that GRS~1915+105's variability states could be classified into three distinct classes: chaotic, stochastic, and ``nonstochastic'' (chaotic + noise).
In a follow-up work, \cite{harikrishnan11} supplemented prior studies with further detailed analysis, which revealed some of the challenges in applying nonlinear time series analysis to light curves corrupted with noise.
To address some of these concerns, \cite{harikrishnan11} recommended further investigations.
More recent explorations by \cite{sukova16} and \cite{jacob16} have also presented results hinting at the presence of determinism in some of the variability states.
In contrast to the suggestions in the aforementioned studies, after careful investigation and taking into consideration several pitfalls, we find no evidence of low-dimensional chaos or determinism in any of the GRS~1915+105 light curves we investigated.
Many of the light curves of GRS~1915+105 obtained from \emph{RXTE} measurements are short and nonstationary,
apart from being corrupted with variegated forms of noise.
Therefore, until long stationary light curves of GRS~1915+105 are available, classification strategies based on nonlinear measures offer little or no advantage over other schemes and may lead to misleading results.
These are the main results of this paper.
In the rest of the paper, we exhaustively investigate the question in hand to reach our conclusions.


\section{Nonlinear time series analysis}

In this section, we briefly present the techniques of nonlinear time series analysis used in this paper.
More detailed discussions, including potential pitfalls in using these methods, are presented later in the appendices.


\subsection{Phase space reconstruction}

A light curve exists in the form of a discrete time series $\{x_i\}=\{x_1, x_2, \ldots x_N\}$.
The typical first step in nonlinear time series analysis is to use these measurements to form $d$-dimensional delay vectors,
\begin{equation}
  \mathbf{y}_i^{(d)} = (x_i, x_{i + \tau}, x_{i + 2\tau}, \ldots, x_{i + (d - 1)\tau}),
\end{equation}
where $d$ is called the embedding dimension, $\tau$ is an integer called the time delay, and the index $i$ runs from $1$ to $M=N-(d-1)\tau$.
For a time series of length $N$, this results in a $d$-dimensional time series $\{\mathbf{y}^{(d)}_i\}$ of length $M$.
Suppose the original time series $\{x_i\}$ is the output of a chaotic system that has a fractal attractor of box dimension $D_0$.
In such a case, by virtue of Takens's embedding theorem \citep[or rather through its extension by \citealt{sauer91}]{takens81} and related ideas \citep{packard80, aeyels81, mane81}, it is guaranteed under generic conditions that the phase space formed by the delay vectors is a diffeomorphically equivalent reconstruction of the original attractor whenever $d > 2D_0$.
Thus, the invariants used to characterize the original chaotic system can be estimated by calculating them on the reconstructed attractor.
On the other hand, for a stochastic time series, the delay vectors do not settle down on an attractor and fill up the phase space instead.

For successful reconstruction, one has to carefully choose the values of $d$ and $\tau$.
Since we generally do not know anything about the original system beforehand, including the value of $D_0$, one cannot simply take $d > 2D_0$.
Furthermore, since $d > 2D_0$ is only a sufficient condition and not a necessary one, it is possible that there is a minimum embedding dimension $d_{\min} < 2D_0$ at which the reconstruction is successful.
Though we revisit this point in the next section, we remark that one strategy for selecting an appropriate $d_{\min}$ is to follow the dependence of the values of the computed invariants with increasing $d$.
The saturation in the value of the invariants marks the minimum $d$.
As far as the time delay is concerned, in this paper we take it to be either the first zero-crossing time of the autocorrelation function or the autocorrelation time (i.e., the $1/e$ decay time of the autocorrelation function) of the time series. (See also Appendix~\ref{app:time-delay}.)


\subsection{Dimension and entropy estimates}


\subsubsection{Correlation dimension}
\label{sec:d2}

A widely used estimate of the fractal dimension is the correlation dimension introduced by \cite{grassberger83a}.
To estimate the correlation dimension of a fractal attractor from a measured time series, we begin by computing the correlation sum $C_2^{(d)}(r)$ using the delay vectors $\mathbf{y}_i^{(d)}$:
\begin{equation}
  \label{eq:corrsum}
  C_2^{(d)}(r) = \frac{\sum_{i = 1}^{M - 1} \sum_{j = i + 1}^{M} \Theta (r - \|\mathbf{y}_i^{(d)} - \mathbf{y}_j^{(d)}\|)}{\frac{1}{2} M(M - 1)}.
\end{equation}
Here $M = N - (d - 1)\tau$ is the total number of points in the reconstructed $d$-dimensional phase space and $\Theta$ is the Heaviside step function: $\Theta(r) = 1$ for $r > 0$ and $\Theta(r) = 0$ otherwise.
$C_2^{(d)}(r)$ can be interpreted as the fraction of point pairs in the phase space that are closer than a distance $r$.
For a fractal attractor, in the limit $r \rightarrow 0$, $C_2^{(d)}(r)$ scales as $r^{D_2}$ whenever $d > D_2$ \citep{ding93}.
The scaling exponent $D_2$, which is a positive real number, is called the correlation dimension.

Since $D_2$ is not known beforehand, in practical calculations, beginning from an embedding dimension of $d = 1$, one would examine the plot of the correlation sum $C_2^{(d)}(r)$ for a scaling region and compute $D_2(d)$ in the region.
Instead of directly inspecting the correlation sum plots, it is often customary to plot the local slope,
\begin{equation}
  \label{eq:d2r}
  D_2^{(d)}(r) = \frac{\mathrm{d}\log{C_2^{(d)}(r)}}{\mathrm{d}\log{r}},
\end{equation}
as a function of $r$ in order to identify this scaling region.
Scaling at extreme length scales is often impaired by `edge effects,' i.e., at very small length scales, noise and poor statistics create fluctuations in $D_2^{(d)}(r)$, and at very large length scales, finiteness of attractor size causes it to drop to zero.
A scaling region with a fairly constant $D_2(d)$ is generally found at intermediate length scales sandwiched between these two extremes.
If the time series comes out of a low-dimensional chaotic process, then one expects $D_2(d)$ to converge to the $D_2$ of the underlying attractor within a sufficiently small $d$ (i.e., at the minimum embedding dimension).
On the other hand, for stochastic processes, the delay vectors fill up the phase space, causing $D_2(d)$ to diverge with $d$.
For the special case of uncorrelated noise, $D_2(d)$ would be equal to $d$.

Correlation dimension quantifies the spatial clustering of points on the attractor and does not contain any dynamical or temporal information.
Thus, ideally, one should draw an independent sample of point pairs $(i, j)$ from the attractor when computing the correlation sum using Equation~(\ref{eq:corrsum}).
However, the delay vectors $\mathbf{y}^{(d)}_i$ obtained from a time series of finite length cannot be considered as statistically independent since each point is correlated to its immediate neighbors owing to continuous evolution in time.
Such temporal correlations between points, different from spatial correlations due to the immanent geometry of the attractor, often result in a spurious lowering of the computed $D_2$.
\cite{theiler86} showed that this problem can be avoided by excluding point pairs that have a temporal separation less than a fixed threshold $W$, often called the Theiler window \citep{galka00,kantz04}.
Thus, instead of using Equation~(\ref{eq:corrsum}), one should use
\begin{equation}
  \label{eq:corrsum-theiler}
  C_2^{(d)}(r) = \frac{\sum_{i = 1}^{M - W - 1} \sum_{j = i + W + 1}^{M} \Theta (r - \|\mathbf{y}_i^{(d)} - \mathbf{y}_j^{(d)}\|)}{\frac{1}{2}(M - W)(M - W - 1)}
\end{equation}
to calculate the correlation sum.
Since we lose only about $O(WM)$ pairs of the original $O(M^2)$ pairs, deterioration in statistics due to this alteration is negligible.
\cite{grassberger90} recommends us ``to be very generous'' with the choice of the Theiler window $W$.
For a truly fractal attractor, the estimated dimension is independent of the choice of $W$, and tends to converge rapidly with increasing $W$.
As a conservative estimate for $W$, one may use the autocorrelation time of the scalar time series $\{x_i\}$.

Since the quality of dimension estimates also depends on the length of the time series, several suggestions for estimating the minimum required lengths have been made (see \citealt{tsonis92} for a review).
A frequently used estimate for the required length is the one given by \cite{eckmann92} who showed that given a time series of length $N$, dimension estimates larger than $2\log_{10}{N}$ cannot be justified.
Thus, after estimating $D_2$, one should additionally verify that it satisfies this criterion.


\subsubsection{Correlation entropy}

\cite{grassberger83} showed that a close lower bound to the Kolmogorov--Sinai entropy, called the correlation entropy $K_2$ can be obtained directly from the correlation sum (Equation~\ref{eq:corrsum-theiler}):
\begin{equation}
  K_2 = \lim_{d \rightarrow \infty} \lim_{r \rightarrow 0} K_2^{(d)}(r),
\end{equation}
where $K_2^{(d)}(r)$ is
\begin{equation}
  K_2^{(d)}(r) = \frac{1}{\tau}\log{\frac{C_2^{(d)}(r)}{C_2^{(d + 1)}(r)}}.
\end{equation}
Since experimental data is always of finite length and resolution, the limits $d \rightarrow \infty$ and $r \rightarrow 0$ can never be achieved in practice.
For small values of $r$, contamination due to noise would become prominent, and the entropy increases.
At large values of $r$, scaling is again broken due to the finite size of the attractor, and one would typically measure smaller values of the entropy.
With finite data, one would look for a plateau in the plot of $K_2^{(d)}(r)$ versus $r$ where $K_2^{(d)}(r)$ is relatively independent of $r$.
By extrapolating the value of $K_2(d)$ computed on this plateau with $d$, we can obtain an estimate of $K_2$.
For regular deterministic dynamics (e.g., a periodic orbit) $K_2$ is equal to zero, whereas it is nonzero and positive for chaotic systems and infinite for stochastic systems.
Thus, an estimate of the $K_2$ entropy, like $D_2$, would also help in the positive identification of chaos.
However, due to the difficulty in obtaining reliable estimates of the $K_2$ entropy from short and noisy data sets, it has remained less popular than $D_2$ in practical nonlinear time series analysis \citep{galka00}.


\subsection{False nearest neighbors}
\label{sec:false-nearest-neighbors}

As a final test for determinism, we consider the method of false nearest neighbors, introduced by \cite{kennel92}.
It was originally developed as an alternative to the approach of observing the saturation in invariants to determine the minimum embedding dimension required to reconstruct an attractor.
If the attractor has not unfolded properly, it will contain a large number of `false nearest neighbors,' i.e., points that are close together solely due to trajectory crossings caused by the projection of the attractor onto a phase space of smaller dimension.
Such neighbors no longer remain close once we increase the embedding dimension, telling us that the attractor cannot be fully unfolded at the chosen embedding dimension.
Operationally, we consider a point $\mathbf{y}_i^{(d)}$ and its nearest neighbor $\mathbf{y}_{n(i, d)}^{(d)}$ in the $d$-dimensional reconstructed phase space as false if
\begin{equation}
  \label{eq:fnn1}
  \frac{|x_{i + d\tau} - x_{n(i, d) + d\tau}|}{\| \mathbf{y}_i^{(d)} - \mathbf{y}_{n(i, d)}^{(d)} \|} > A,
\end{equation}
where $\| \cdot \|$ is the Euclidean norm and $A$ is a suitable threshold.
We test this condition for all points in the phase space and compute the total fraction of false near neighbors for each $d$.
If this fraction becomes zero at a finite, and preferably small $d$ (the minimum embedding dimension $d_{\min}$), then we can conclude that the attractor has properly unfolded and that the time series comes out of a deterministic process.
In other words, the system generating the time series is low-dimensional enough to allow its attractor to be embedded in a phase space of $d_{\min}$ dimensions.

In the case of a stochastic time series, which is infinite dimensional, we expect the fraction of false neighbors to never become zero.
Nonetheless, short noisy time series for which near neighbor distances, $\| \mathbf{y}_i^{(d)} - \mathbf{y}_{n(i, d)}^{(d)} \|$, increase unboundedly with $d$ may also appear to have no false neighbors.
As a workaround for this caveat, in addition to the first test (Equation~\ref{eq:fnn1}) a second test is introduced, and two near neighbors are also declared as false if
\begin{equation}
  \label{eq:fnn2}
  \frac{\| \mathbf{y}_i^{(d + 1)} - \mathbf{y}_{n(i, d)}^{(d + 1)} \|}{\sigma} > B,
\end{equation}
where $\sigma$ is the standard deviation of the scalar time series $\{x_i\}$ and $B$ is another threshold.
Although this procedure is reasonably independent of the values we choose for the threshold parameters, in our analysis we use $A = 10.0$ and $B = 2.0$ following the recommendations of \cite{kennel92}.


\subsection{Surrogate analysis}
\label{sec:surrogates}

Tests for deterministic chaos, for various reasons (see Appendix~\ref{app:potential-pitfalls}), can often produce spurious results for time series that are not deterministic.
Though identifying the presence of nonlinearity in a time series is a less ambitious goal than identifying chaos, it is less error-prone and easier.
Nevertheless, since the presence of chaos requires nonlinearity, in cases where there is ambiguity in the results, it makes sense to test for nonlinearity as well.

\cite{theiler92} introduced surrogate analysis as a statistical technique for detecting nonlinearity in time series.
In surrogate analysis, instead of attributing the observed properties of a time series to nonlinearity, we assume a null hypothesis that our results are consistent with a specific stationary linear stochastic process.
We then generate an ensemble of surrogate data sets that conform to the chosen null hypothesis and the significance level one wishes the test to have.
One then computes a nonlinear discriminating parameter (e.g., an estimate of $D_2$, fraction of false neighbors, etc.) using the original and the surrogate data sets.
If the differences in the computed parameters for the original and the surrogates are significantly different, we reject the null hypothesis and conclude that it is insufficient to explain our results.
In this sense, null hypothesis testing is the statistical analog of proof by contradiction.
For surrogate analysis of the GRS~1915+105 light curves, we consider three null hypotheses:
\begin{figure*}
  \begin{center}
    \includegraphics[width=\textwidth]{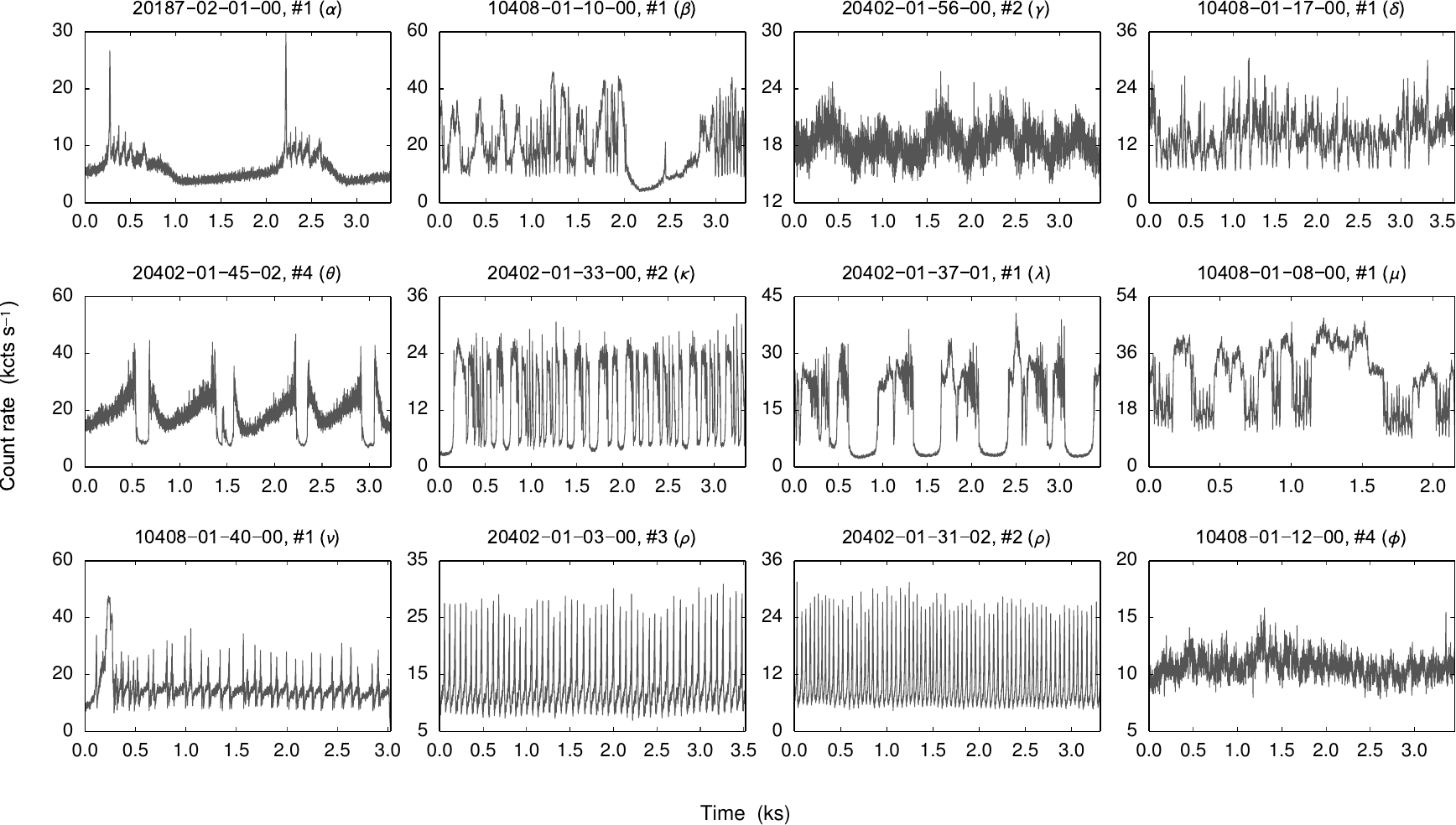}
  \end{center}
  \caption{
    Sample light curves representing the different GRS~1915+105 variability classes that we study in this paper.
    The light curves have a temporal resolution of $0.5$s and a maximum observation time of $\sim 3600$s.
    The \emph{entire} light curve corresponding to a specific observation and interval (indicated by \#) is shown.
    Note the strong lack of stationarity in most light curves.\label{fig:curves}
  }
\end{figure*}

\paragraph{Null Hypothesis NH1} A rudimentary null hypothesis (NH1) could be that the time series is composed of uncorrelated random numbers (i.e., white noise) drawn from an unknown but stationary distribution.
A random permutation of the original time series is a valid surrogate series under NH1.
A rejection of NH1 would indicate that the points in the time series are correlated in some way (linearly or nonlinearly).

\paragraph{Null Hypothesis NH2} A broader and more realistic null hypothesis (NH2) is that the time series is the output of a stationary linear stochastic process with Gaussian inputs that has possibly gone through a static monotonic transformation during measurement.
Considering the broadness of NH2, a rejection would indicate the presence of nonlinearity in the time series.
Noise with the same linear correlations and probability distribution as the original time series would be the appropriate surrogate series for testing NH2 \citep{schreiber96}.
If NH2 cannot be rejected, one may additionally test NH1 to check whether the time series is correlated at all.

\paragraph{Null Hypothesis NH3} For time series with periodic components, the null hypothesis of a linear stochastic process is not appropriate, as one would anyway not expect such a process to be periodic.
Due to the omnipresence of dissipation and nonlinearity in nature, periodic signals are traditionally modeled as limit cycle solutions of a nonlinear dynamical system.
In fact, nonlinear systems can exhibit robust periodic oscillations through limit cycles even in the presence of noise \citep{kantz04,samanta14}.
For such time series, \cite{theiler94} proposed a more plausible hypothesis (NH3): the time series is the output of a periodic nonlinear oscillator corrupted with observational and/or dynamical noise of some kind, with no dynamical correlation between individual periodic cycles.
For testing NH3, individual cycles in the original time series are scrambled randomly and spliced together to create \emph{cycle-shuffled surrogates}.
Since the original time series is periodic, its cycle-shuffled surrogates are also visually periodic, though they do not carry any dynamical correlation beyond the time scale of an average oscillation.
A rejection of NH3 would suggest the presence of dynamical correlations between the periodic cycles, i.e., a particular cycle affects the appearance of the one following it.

\vfill


\section{Data, results, and discussion}
\label{sec:data-results-discussion}

Having discussed the theoretical ideas behind the numerical techniques, we are now ready to begin our technical investigation of the X-ray light curves of GRS~1915+105.
We analyze light curves representing the 12 different variability classes of GRS~1915+105 \citep{belloni00} using data from the \emph{RXTE} Target of Opportunity archive.\footnote{\href{https://heasarc.gsfc.nasa.gov/docs/xte/xtegof.html}{https://heasarc.gsfc.nasa.gov/docs/xte/xtegof.html}}
The data collected by \emph{RXTE} is not continuous due to the Earth's occultation, and hence each observation consists of one or more continuous intervals separated by gaps.
The maximum length of an interval is roughly 1 hour.
For concreteness, we choose some specific observations and intervals, as detailed in Table~\ref{tab:results-summary}, with a motivation of contrasting our results with existing studies in the literature.
We bin the light curves at a time resolution of $0.5$s in the $0$--$35$ proportional counter array channel range (corresponding to an energy range of $\sim$ $2$--$15$~keV).
This results in about $4000$--$7200$ points in each of the light curves that we use in our study.
The choice of $0.5$s is a good balance between oversampling and the required time resolution for practical time series analysis.
Some examples of the light curves we analyze in this paper are shown in Figure~\ref{fig:curves}.
\begin{deluxetable*}{ccccccccc}
  \tablewidth{0pt}
  \tablecaption{Summary of results for light curves from the 12 GRS~1915+105 classes\label{tab:results-summary}}

  \tablehead{
    \colhead{Observation ID}                                         &
    \colhead{Interval}                                               &
    \colhead{Class}                                                  &
    \colhead{Length $N$}                                             &
    \colhead{Stationarity\tablenotemark{a}}                          &
    \colhead{Time Delay $\tau$\tablenotemark{b}}                     &
    \multicolumn{3}{c}{Surrogate Analysis\tablenotemark{c}}          \\
    \cline{7-9} \\
    &  &  &  &  &  &
    \colhead{NH1}                                                    &
    \colhead{NH2}                                                    &
    \colhead{NH3}
  }
  \startdata
  20187-02-01-00    &    1    &     $\alpha$    &    6615    &    NS    &    479    &    \nodata    &     R    &     \nodata \\
  20187-02-01-00    &    2    &     $\alpha$    &    5250    &    NS    &    539    &    \nodata    &     R    &     \nodata \\
  10408-01-10-00    &    1    &     $\beta$     &    6400    &    NS    &     94    &    \nodata    &     R    &     \nodata \\
  20402-01-53-00    &    2    &     $\beta$     &    6480    &    NS    &    300    &    \nodata    &     R    &     \nodata \\
  20402-01-56-00    &    2    &     $\gamma$    &    6600    &     S    &     7     &    \nodata    &     R    &     \nodata \\
  20402-01-39-00    &    1    &     $\gamma$    &    6272    &     S    &     4     &    \nodata    &     R    &     \nodata \\
  10408-01-17-00    &    1    &     $\delta$    &    7200    &    NS    &     22    &    \nodata    &     R    &     \nodata \\
  10408-01-18-04    &    1    &     $\delta$    &    4802    &     S    &     27    &       R       &    NR    &     \nodata \\
  20402-01-45-02    &    4    &     $\theta$    &    6174    &    NS    &     78    &    \nodata    &     R    &     \nodata \\
  10408-01-15-00    &    1    &     $\theta$    &    4860    &    NS    &     88    &    \nodata    &     R    &     \nodata \\
  20402-01-33-00    &    2    &     $\kappa$    &    6400    &     S    &     35    &    \nodata    &     R    &     \nodata \\
  20402-01-35-00    &    1    &     $\kappa$    &    6615    &     S    &     18    &    \nodata    &     R    &     \nodata \\
  20402-01-37-01    &    1    &    $\lambda$    &    6615    &    NS    &    173    &    \nodata    &     R    &     \nodata \\
  10408-01-38-00    &    3    &    $\lambda$    &    6480    &    NS    &    295    &    \nodata    &     R    &     \nodata \\
  10408-01-08-00    &    1    &      $\mu$      &    4235    &    NS    &    119    &    \nodata    &     R    &     \nodata \\
  20402-01-53-01    &    1    &      $\mu$      &    6272    &     S    &     6     &    \nodata    &     R    &     \nodata \\
  10408-01-40-00    &    1    &      $\nu$      &    5775    &    NS    &     54    &    \nodata    &     R    &     \nodata \\
  10408-01-44-00    &    4    &      $\nu$      &    3780    &    NS    &    336    &    \nodata    &     R    &     \nodata \\
  20402-01-03-00    &    3    &      $\rho$     &    6976    &     S    &     21    &    \nodata    &     R    &     R       \\
  20402-01-31-02    &    2    &      $\rho$     &    6490    &    NS    &     17    &    \nodata    &     R    &     R       \\
  10408-01-12-00    &    4    &      $\phi$     &    6615    &    NS    &     29    &       R       &    NR    &     \nodata \\
  10408-01-19-00    &    1    &      $\phi$     &    4125    &     S    &     60    &       R       &    NR    &     \nodata \\
  20402-01-05-00    &    3    &      $\chi$     &    5670    &     S    &     1     &       R       &    NR    &     \nodata \\
  10408-01-22-00    &    2    &      $\chi$     &    3872    &     S    &     1     &       R       &    NR    &     \nodata \\
  \enddata
  \tablecomments{
    No evidence of chaos or determinism is seen in any class.
    Though we can reject the null hypothesis of a linear stochastic process with Gaussian inputs (NH2) for several classes, many of the light curves are also nonstationary, which could be the source of rejection rather than nonlinearity.
  }
  \tablenotetext{a}{Stationarity (S) or nonstationarity (NS) of the light curves as reported by the chi-squared test for stationarity at a significance level of $0.05$ (Appendix~\ref{app:stationarity-test}).}
  \tablenotetext{b}{Time delay $\tau$ is equal to the autocorrelation time for light curves from all classes except $\rho$, for which it is equal to the first zero-crossing time of the autocorrelation function (Appendix~\ref{app:time-delay}).}
  \tablenotetext{c}{Rejection (R) or nonrejection (NR) of null hypotheses representing uncorrelated random numbers or white noise (NH1), correlated Gaussian noise possibly transformed by a static monotonic function (NH2), and a noisy periodic orbit (NH3).  NH1 needs to be tested only in cases where NH2 cannot be rejected.  NH1, NH2, and NH3 have been rejected at significance levels of $0.01$, $0.05$, and $0.05$, respectively.}
\end{deluxetable*}

As the first step in our analysis, we investigate the stationarity of the light curves.
Using a chi-squared test of stationarity (Appendix~\ref{app:stationarity-test}), we can reject the null hypothesis of stationarity at a significance level of $0.05$ for more than half the light curves~(Table~\ref{tab:results-summary}).
A visual inspection of the light curves also agrees with these results intuitively (see Figure~\ref{fig:curves}).
The weak nonstationarity of certain light curves, such as those from class $\rho$, may be due to the presence of observational noise.
On the other hand, for a majority of the light curves, lone bursts and dips that stand out are the primary sources of nonstationarity.
In principle, this may be due to the inherent nonstationarity in the physical processes behind these light curves themselves.
However, since the light curves that we analyze are finite in length, stationarity (or nonstationarity) as indicated by the chi-squared test only applies to the light curve obtained during the specific interval of observation, rather than the whole class.
Furthermore, considering how the different variability classes frequently reappear in observations, it is more likely that this nonstationarity is due to the short intervals of continuous observation (inevitable because of interruptions caused by the Earth's occultations).
In any case, the best we can do is to interpret our results in light of this apparent nonstationarity.

Next, we estimate the time delay $\tau$ required for constructing the delay vectors (see Appendix~\ref{app:time-delay} for more details).
We do not use time-delayed mutual information to estimate $\tau$ since it does not have clear minima for any of the light curves.
The $1/e$ decay time of the autocorrelation function (i.e., the autocorrelation time) is indicative of the average time for $\{x_{i}\}$ and $\{x_{i+\tau}\}$ to get linearly decorrelated.
In other words, the $\tau$ we choose must at least be of the order of the autocorrelation time to ensure average general independence between the components of the delay vector.
Thus, for light curves from classes other than $\rho$, we choose the autocorrelation time as the time delay.
For the two light curves from class $\rho$, both of which have periodic properties, we use a $\tau$ equal to the first zero-crossing time of the autocorrelation function (which is slightly larger than the corresponding autcorrelation time).
The values of $\tau$ we use for each light curve are tabulated in Table~\ref{tab:results-summary}.
If the chosen $\tau$ is too small, most points of the reconstructed attractor would be concentrated near the phase space diagonal, which can often lead to spurious results.
Thus, to ensure that the $\tau$ we use is sufficiently large, we visually inspect two- and three-dimensional phase portraits of the reconstructed phase space and verify that most points have moved away from the phase space diagonal.
\begin{figure*}
  \begin{center}
    \includegraphics[width=\textwidth]{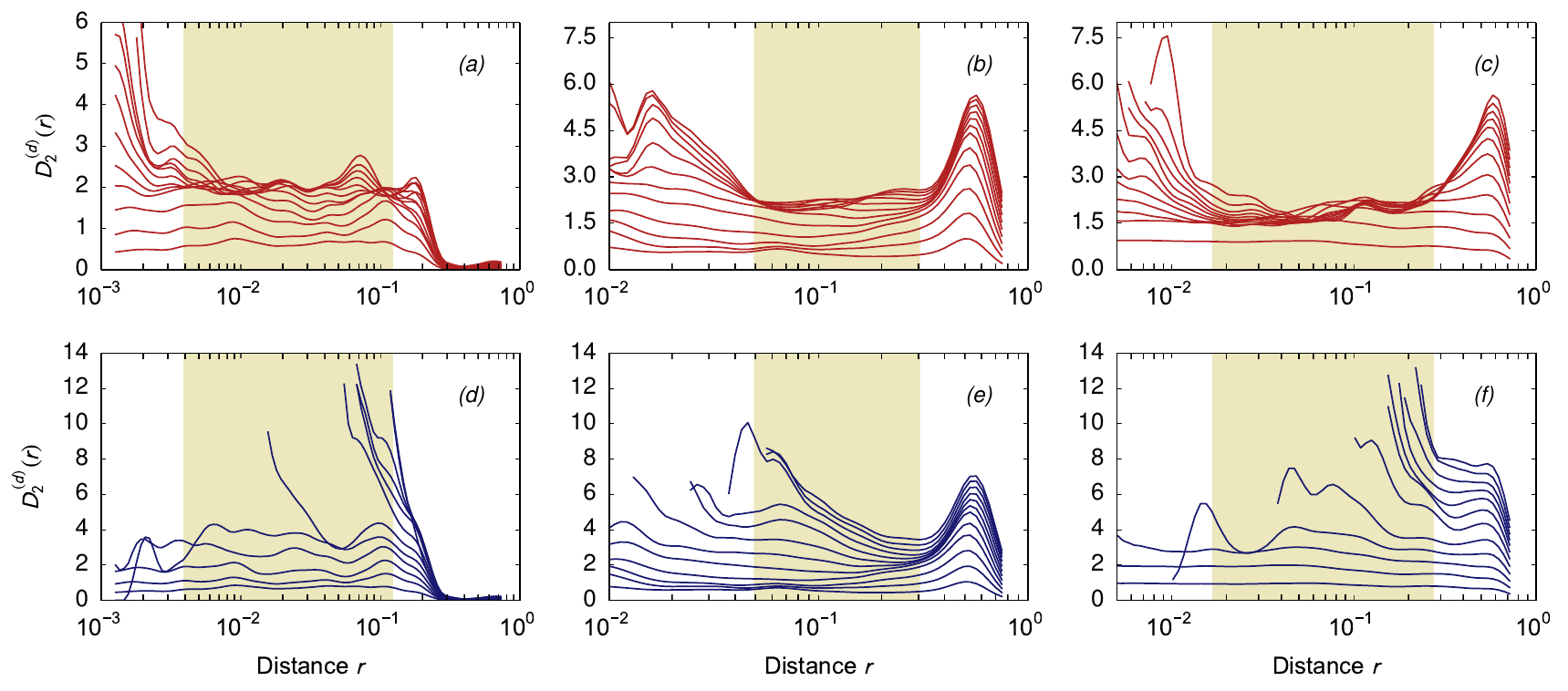}
  \end{center}
  \caption{
    Local correlation dimension $D^{(d)}_2(r)$ as a function of the distance $r$ at embedding dimensions $d = 1, 2, \ldots, 12$ (counting from below) for (a), (d) class $\alpha$, observation 20187-02-01-00, 1st interval, (b), (e) class $\lambda$, observation 20402-01-37-01, 1st interval, and (c), (f) class $\mu$, observation 10408-01-08-00, 1st interval.
    No Theiler window is used in computing $D^{(d)}_2(r)$ in the upper panels, whereas a Theiler window equal to the autocorrelation time of each light curve is used in the lower panels.\label{fig:d2}
    Scaling regions chosen to avoid edge effects are highlighted in yellow.
  }
\end{figure*}

Before proceeding to compute the correlation sum, we linearly rescale the light curves to the interval $[0, 1]$ to make comparisons between the different variability classes easier and to bring about a uniformity in the analysis.
After constructing the delay vectors, we compute the interpoint distances using the Chebyshev (maximum norm) metric and vary the embedding dimension between $d = 1$ and $d = 12$ for all light curves, except that of class $\alpha$, observation 20187-02-01-00, 2nd interval for which the largest embedding dimension is $d = 10$.
We first use Equation~(\ref{eq:corrsum}) to compute the correlation sum without employing any Theiler window.
We then find the local correlation dimension $D_2^{(d)}(r)$ (Equation~\ref{eq:d2r}) by performing a local least-squares fit \citep{galka00} on the slopes of the $C_2^{(d)}(r)$ curves, which suppresses fluctuations due to noise and shows the variations in scaling clearly (some examples are shown in Figure~\ref{fig:d2}).
We observe scaling regions in the $D_2^{(d)}(r)$ curves for light curves from classes $\alpha$, $\beta$, $\delta$, $\kappa$, $\lambda$, and $\mu$ for all $d$, with the value of $D_2(d)$ measured in the scaling region converging after about $d = 4$.
The converged values of $D_2$ for these classes vary between $\approx 2.0$ and $4.5$.
Note that since the shortest light curve we use in our analysis has about $N \approx 4200$ points, the above values of $D_2$ pass the consistency criterion proposed by \cite{eckmann92}, according to which estimates of $D_2$ should not be larger than $2\log_{10}N \approx 7.2$.
For classes $\theta$ and $\nu$, only light curves from observations 10408-01-15-00 and 10408-01-40-00 show a clear scaling range and a limiting value of $D_2(d)$.
Though clear intermediate scaling regions are visible for the light curves of class $\rho$, we do not observe a saturation in the value of $D_2(d)$.
For the light curves of classes $\gamma$, $\phi$, and $\chi$, we observe scaling at small to intermediate distances, and the local correlation dimension often goes as high as the embedding dimension, without any saturation.
\begin{figure}
  \begin{center}
    \includegraphics[width=\columnwidth]{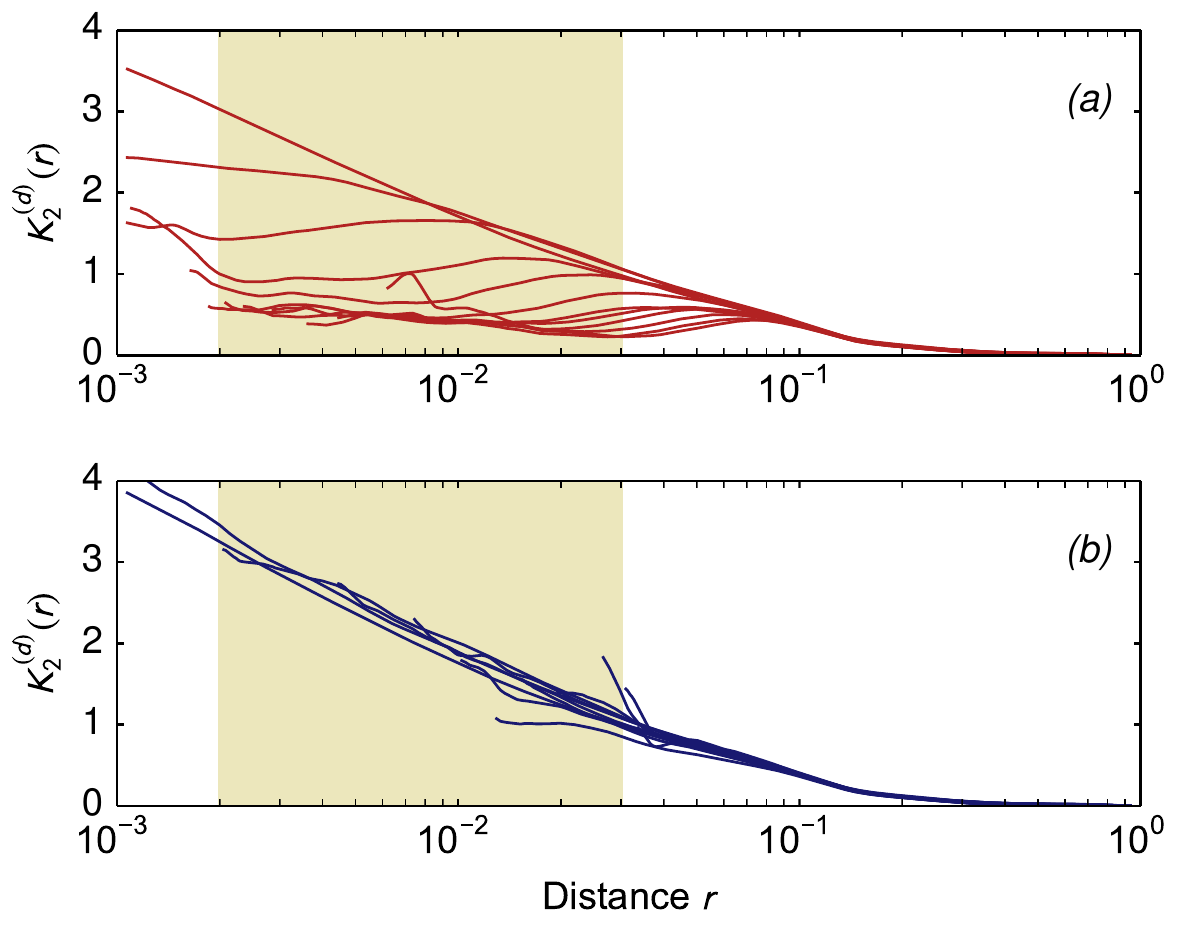}
  \end{center}
  \caption{
    Local correlation entropy $K^{(d)}_2(r)$ as a function of the distance $r$ for class $\nu$, observation 10408-01-40-00, 1st interval, for embedding dimensions $d = 1, 2, \ldots, 12$ (counting from above in (a)), computed using a Theiler window (a) $W = 0$ and (b) $W = 54$.
    Note the disappearance of the scaling region in (b).
    $K^{(d)}_2(r)$ is given in units of the time delay $\tau$.\label{fig:entropy}
    The scaling region chosen to avoid edge effects is highlighted in yellow.
  }
\end{figure}

\begin{figure*}
  \begin{center}
    \includegraphics[width=\textwidth]{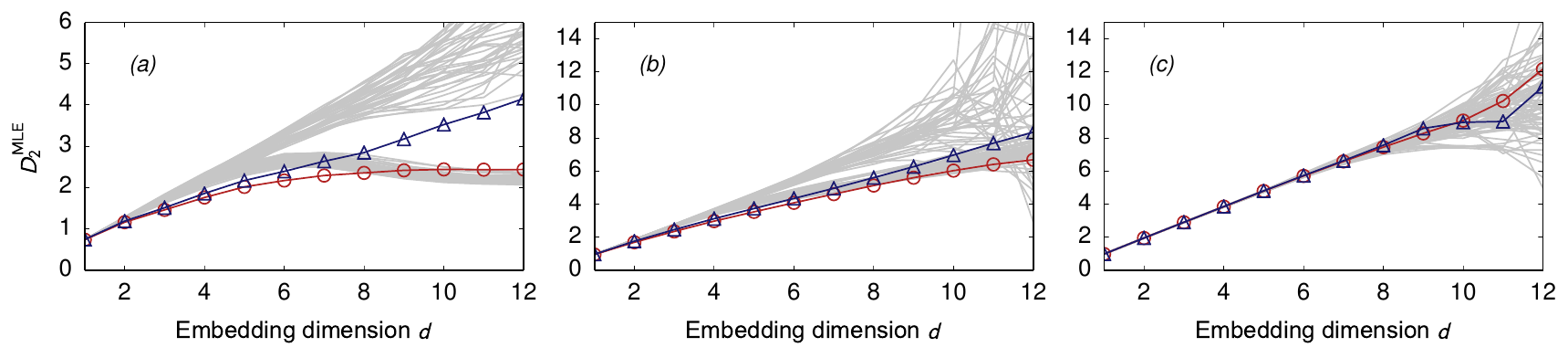}
  \end{center}
  \caption{
    Takens's maximum likelihood estimate $D_2^{\text{MLE}}$ (Appendix~\ref{app:takens-estimator}) as a function of the embedding dimension $d$, computed without using a Theiler window (red with circles) and using a Theiler window equal to the autocorrelation time of the light curves (blue with triangles) for (a) class $\kappa$, observation 20402-01-33-00, 2nd interval (b) class $\theta$, observation 20402-01-45-02, 4th interval, and (c) class $\phi$, observation 1408-01-12-00, 4th interval.
    $D_2^{\text{MLE}}$ has been evaluated at an $r_{{\max}}$ equal to one-half of the standard deviation of the light curves, and thus it includes the scaling information at all distances below $r_{{\max}}$.
    The upper and the lower gray curves in the background show the corresponding $D_2^{\text{MLE}}$ for the 39 surrogate series representing null hypothesis NH2, which is rejected only by classes $\kappa$ and $\theta$.\label{fig:ttmle}
  }
\end{figure*}

However, on repeating our calculations using a Theiler window $W$ equal to the autocorrelation time\footnote{In other words, the $W$ we use is numerically equal to the embedding delay $\tau$ for all light curves except those from class $\rho$, in which case the chosen $W$ is slightly smaller than $\tau$ ($W = 13$ and $12$ for observations 20402-01-03-00 and 20402-01-31-02, respectively).} of the light curves, we observe no saturation in $D_2{(d)}$ with increasing embedding dimension for any of the classes.
For illustrative purpose, $D_2^{(d)}(r)$ curves for a selected set of light curves with and without imposing a Theiler window are shown in Figure~\ref{fig:d2}.
In certain cases, such as class $\alpha$, the scaling at higher dimensions is broken and we get much higher values for the local $D_2^{(d)}(r)$.
We will try to explain these results.
As we have noted previously, more than half the light curves of GRS~1915+105 are nonstationary and have strong temporal correlations, as indicated by their large autocorrelation times.
Employing a Theiler window $W$ results in the removal of $O(WM)$ temporally correlated pairs from a phase space originally containing $O(M^2)$ point pairs.
Had the light curves of GRS~1915+105 been long and stationary enough, a large number of close neighbors would have been present in the vicinity of most points that would have maintained the scaling, irrespective of the value of $W$.
In the case of short and highly nonstationary light curves such as those from classes $\alpha$, $\beta$, $\lambda$, $\nu$, etc., all of which have large autocorrelation times, most points are temporally correlated.
Introducing a Theiler window equal to the autocorrelation time while analyzing these light curves would leave us with very few points to analyze, thus breaking the scaling in $D_2(r)$.
However, even for the reasonably long and stationary light curves of class $\kappa$ the saturation in $D_2(d)$ is lost with a nonzero Theiler window [Figure~\ref{fig:ttmle}(a)].
Considering all this, the convergences in $D_2{(d)}$ observed earlier should be interpreted as being caused by temporal correlations, rather than evidence of low-dimensional chaos (Appendix \ref{app:dimension-entropy-estimates}).
The results for classes $\gamma$, $\rho$, $\phi$, and $\chi$, for which we do not observe any saturation in $D_2(d)$, remain qualitatively the same, but with higher estimates for the local $D_2^{(d)}(r)$.
Though correlation dimension analysis has essentially failed to find evidence of chaos in any of the light curves, it has helped us to clarify many of the subtleties involved.
\begin{figure*}
  \begin{center}
    \includegraphics[width=\textwidth]{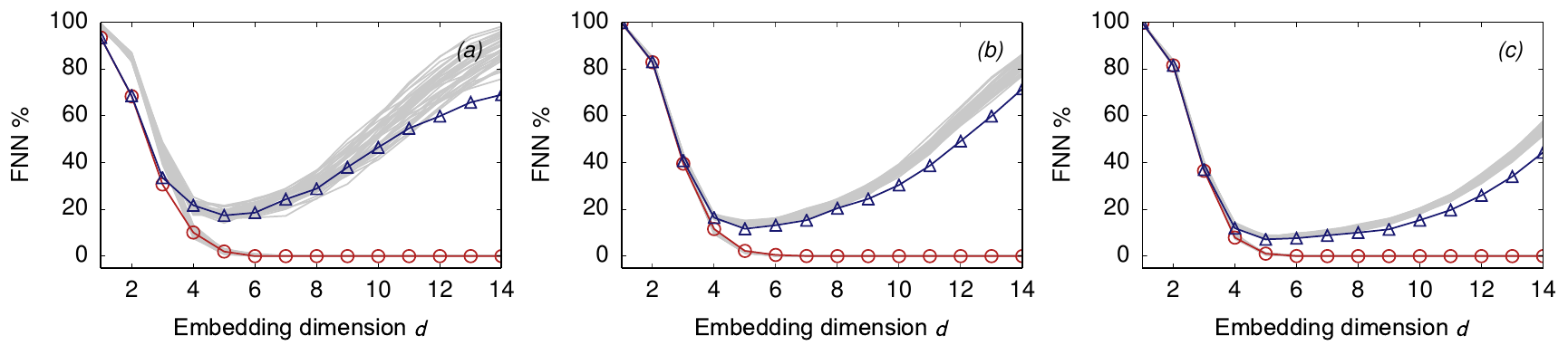}
  \end{center}
  \caption{
    Percentage of false nearest neighbors (FNN) in the reconstructed phase space as a function of the embedding dimension $d$, computed using only the first test (red with circles) computed using both the first and second tests (blue with triangles) for (a) class $\beta$, observation 10408-01-10-00, 1st interval (b) class $\delta$, observation 10408-01-17-00, 1st interval, and (c) class $\gamma$, observation 20402-01-56-00, 2nd interval.
    The upper and lower gray curves in the background show the corresponding FNN fraction for the 39 surrogate series for null hypothesis NH2, which is rejected by all three classes.\label{fig:fnn}
  }
\end{figure*}
\begin{figure*}
  \begin{center}
    \includegraphics[width=\textwidth]{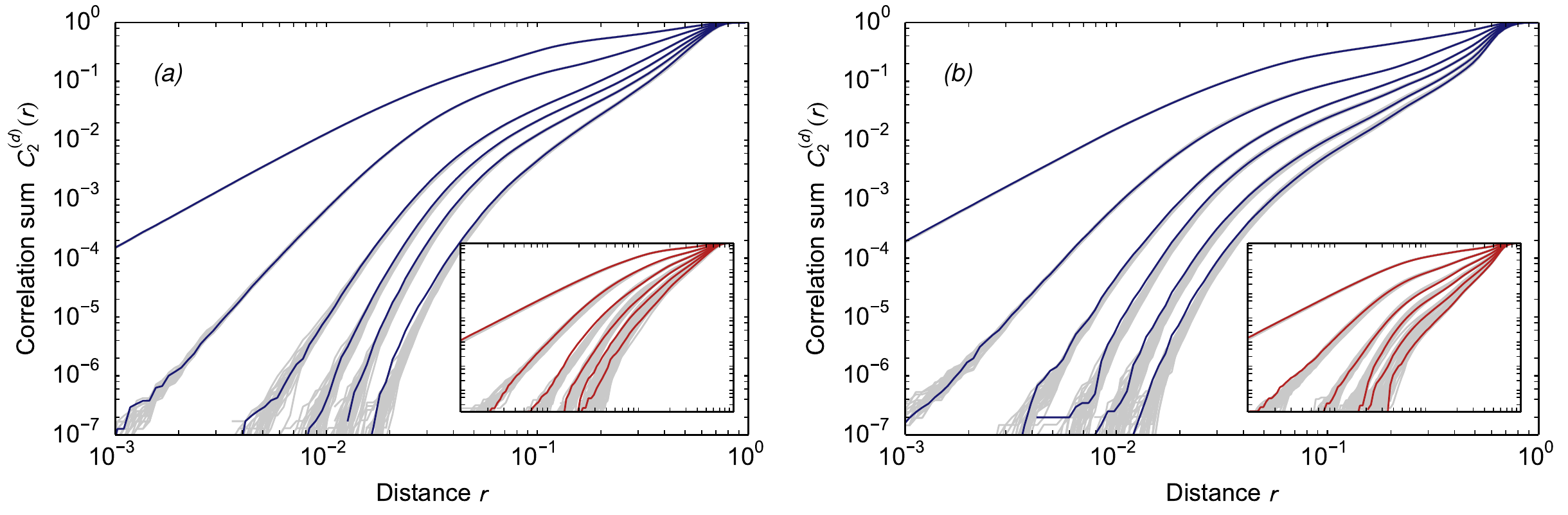}
  \end{center}
  \caption{
    Correlation sum $C_2^{(d)}(r)$ as a function of the distance $r$ at even embedding dimensions $d = 2, 4, \ldots, 12$ (counted from top left) for (a) class $\rho$, observation 20402-01-03-00, 3rd interval and (b) class $\rho$, observation 20402-01-31-02, 2nd interval.
    The gray curves in the background depict the corresponding correlation sums for the 39 cycle-shuffled surrogates satisfying null hypothesis NH3.
    The time delay $\tau$ is the first zero-crossing of the autocorrelation function of the light curves.
    The average time periods of the two light curves are (a) $T \approx 134$ and (b) $T \approx 94$.
    Insets of the figures show $C_2^{(d)}(r)$ for the two light curves computed using a time delay equal to $\tau$ + $T$.
    Though null hypothesis NH3 is technically rejected here, note the close resemblance between the correlation sum curves for the surrogates and the actual light curves.\label{fig:shuffle}
  }
\end{figure*}

As the next step in our analysis, we transform each correlation sum curve to a local correlation entropy $K_2^{(d)}(r)$ curve.
Noise contamination leads to fluctuations in the $C_2^{(d)}(r)$ curves at small distances, and we observe well-defined scaling regions in the $K_2^{(d)}(r)$ curves only for light curves from $\beta$, $\mu$, and $\nu$ classes, with the asymptotic $K_2(d)$ varying between $0.1$ and $0.5$.
($K_2^{(d)}(r)$ for class $\nu$ is shown in Figure~\ref{fig:entropy}.)
The lack of good scaling in the $K^{(d)}_2(r)$ plots is not surprising since computing $K_2$ from $C_2^{(d)}(r)$ is error-prone and subjective, especially with real-world data \citep{galka00}.
However, whatever saturations and scalings seen in the $K^{(d)}_2(r)$ curves are lost once we use a Theiler window equal to the autocorrelation time to compute the correlation sum.
This again shows that the finite values of correlation entropy observed earlier are artifacts of temporal correlations and nonstationarity rather than the presence of chaos.
Thus, the values of $K_2(d)$ that we obtain without using a Theiler window should be interpreted as being caused by the divergence of distant pairs in the reconstructed trajectory rather than due to the divergence of nearby trajectories as would have happened in a chaotic system (Appendix~\ref{app:dimension-entropy-estimates}; \citealt{grassberger91}).

As we remarked earlier, unlike the computation of $D_2$ and $K_2$, the method of false nearest neighbors does not involve the direct estimation of any invariant characteristic of the attractor.
Hence, as a final and alternative attempt to detect determinism, we use this method.
While searching for near neighbors in the reconstructed phase space, we use the Euclidean metric to compute the interpoint distances and impose a minimal temporal separation equal to the autocorrelation time between a point and its neighbor.
Imposing a minimum temporal separation is to prevent the method from incorrectly reporting serially correlated time series, such as $1/f^n$ noise, as low-dimensional (Appendix \ref{app:false-nearest-neighbors}).
The fraction of false near neighbors in the reconstructed phase space computed using only the first test becomes negligible after an embedding dimension of $4$ for light curves from all classes.
Thus, it would appear that these light curves are generated by a low-dimensional chaotic system whose attractor can be embedded in a four-dimensional phase space.
However, this is a misleading result since uncorrelated noise of similar lengths to these light curves would also have zero false neighbors after an embedding dimension of~$4$ when only the first test is used (Appendix~\ref{app:false-nearest-neighbors}).
Since the purpose of the second false near neighbor test is to catch such pitfalls, we now repeat our analysis using both tests.
As can be seen from the illustrative examples in Figure~\ref{fig:fnn}, once both the first and the second tests are used to compute the false neighbor fraction (represented by the upper curves of the figure), we find that it consistently increases with the embedding dimension after $d = 4$ for all the classes.
Thus, once again we are unable ascertain the presence of low-dimensional determinism in any class.

Although we have failed to find conclusive signatures of low-dimensional chaos in any class, it would still be interesting if we can establish the presence of nonlinearity in the light curves through surrogate analysis.
Though the presence of nonlinearity does not say anything about the underlying dimensionality of the system, it indicates that a complex nonlinear process, which could even be a stochastic one, is behind the generation of the light curves (Appendix \ref{app:surrogate-analysis}).
To this end, we generate 39 surrogate series corresponding to null hypothesis NH2 that the given time series is the outcome of a linear stochastic process with Gaussian inputs that has possibly undergone a time-independent monotonic transformation.
These 39 surrogates, along with a two-sided rank-based test, correspond to a significance level of $0.05$ \citep{schreiber00}.
As nonlinear discriminating parameters, we use a maximum likelihood estimate of the correlation dimension (Appendix~\ref{app:takens-estimator}), and the percentage of false nearest neighbors.
Plots of these parameters for the actual light curves, along with their surrogates are shown in Figures~\ref{fig:fnn} and~\ref{fig:ttmle} for a selected set of classes.
This study leads us to reject NH2 for light curves of all classes except $\delta$ (observation 10408-01-18-04), $\phi$, and $\chi$.
As seen from the lower curves of Figure~\ref{fig:ttmle}, we can reject NH2 even in cases where a nonzero Theiler window is not used.
Thus, it would appear that the case for nonlinearity in these light curves is stronger than the case for low-dimensional determinism.
However, among the classes for which we reject NH2, only light curves from $\gamma$, $\kappa$, $\mu$ (observation 20402-01-53-01), and $\rho$ (observation 20402-01-03-00) are stationary.
As surrogate analysis assumes the original time series to be stationary, these rejections could entirely be due to the nonstationarity in the light curves rather than the presence of any nonlinear structure (also see the discussion in Appendix~\ref{app:surrogate-analysis}).
Thus, convincing evidence of nonlinearity is seen only in the light curves of classes $\gamma$, $\kappa$, $\mu$, and $\rho$.
The last three columns of Table~\ref{tab:results-summary} summarize the results of surrogate analysis.

Since we are unable to reject NH2 for classes $\delta$ (observation 10408-01-18-04), $\phi$, and $\chi$, we further attempt to test null hypothesis NH1 that they are uncorrelated random numbers.
Note that we need not test NH1 for classes where we could reject NH2, since a rejection of NH2 anyway indicates the presence of nonlinear correlations.
By comparing the value of the autocorrelation function at unit lag (i.e., $\langle x_{i}x_{i + 1} \rangle$) for these light curves and their 199 randomly permuted surrogates, we reject NH1 at a significance level of $0.01$.

Though it is clear from the rejection of NH2 that class $\rho$ is nonlinear, because of its seeming periodicity, it makes sense to additionally test null hypothesis~NH3, which assumes that it is generated by a periodic nonlinear oscillator with noise of some type.
Note that testing NH3 for other classes would lead to trivial rejections since only light curves from class $\rho$ are periodic.
To this end, we generate 39 cycle-shuffled surrogates and compare the values of $D_2^{\text{MLE}}$ between the actual light curves of class $\rho$ and the surrogates.
We also repeat the analysis using a time delay larger than the average time period of the light curves to prevent potential false acceptances of NH3 owing to limited randomization while generating cycle-shuffled surrogates \citep{small05}.
Although we theoretically reject NH3 for class $\rho$ at a significance level of $0.05$, as seen from Figure~\ref{fig:shuffle}, the correlation sum curves of the cycle-shuffled surrogates closely imitate those of the actual light curves.
Even at a very large $\tau$ equal to the sum of the first zero-crossing of the autocorrelation function of the time series and its average time period, there is good resemblance between the curves.
On the other hand, if class $\rho$ were to have chaotic dynamics, its attractor would have become a noise-like featureless collection of points at such a large time delay (Appendix~\ref{app:time-delay}).
Thus, despite a formal rejection of NH3, the model of a periodic nonlinear oscillator affected by noise of some kind may be considered as a useful approximation for the dynamics of class $\rho$.

Since it has been suggested \citep{misra04,harikrishnan11} that noise contamination in the light curves could be the reason for a lack of saturation in $D_2(d)$ and $K_2(d)$ values, we repeat our analysis after employing a nonlinear noise filter (Appendix~\ref{app:nonlinear-noise-reduction}).
Apart from the fact that this noise reduction scheme is powerful enough to suppress `in-band noise' (i.e., noise with the same power spectra as the light curves), it also leaves the light curve peaks undistorted.
Though minor qualitative improvements are seen in the local $D_2^{(d)}(r)$ and $K_2^{(d)}(r)$ curves after filtering, scalings are not restored, and saturations are not observed in the computed quantities.
In fact, the dimension and entropy estimates showcased in Figures~\ref{fig:d2}--\ref{fig:shuffle} are all obtained using the filtered light curves and their surrogates.
\begin{deluxetable}{cl}
  \tablewidth{0pt}
  \tablecaption{Summary of previous studies on GRS~1915+105 light curves\label{tab:previous-results}}

  \tablehead{Class & Previous results}
  \startdata
    $\alpha$  & Chaotic,\textsuperscript{c} nonstochastic\textsuperscript{a} (chaotic + Poisson noise)                            \\
    $\beta$   & Chaotic,\textsuperscript{{a,b,c}} nonlinear and possibly chaotic\textsuperscript{d}                               \\
    $\gamma$  & Stochastic\textsuperscript{{a,b}} (white noise\textsuperscript{c})                                                \\
    $\delta$  & Nonstochastic\textsuperscript{a} (chaotic + Poisson noise),                                                       \\
              & nonlinear and possibly chaotic,\textsuperscript{d} stochastic\textsuperscript{b} (white noise\textsuperscript{c}) \\
    $\theta$  & Chaotic,\textsuperscript{{b,c,e}} nonlinear and possibly chaotic,\textsuperscript{d}                              \\
              & nonstochastic\textsuperscript{a} (chaotic + Poisson noise)                                                        \\
    $\kappa$  & Chaotic\textsuperscript{{a,b}} (with colored noise\textsuperscript{c}),                                           \\
              & nonlinear and possibly chaotic\textsuperscript{d}                                                                 \\
    $\lambda$ & Chaotic\textsuperscript{{a,b}} (with colored noise\textsuperscript{c}),                                           \\
              & nonlinear and possibly chaotic\textsuperscript{d}                                                                 \\
    $\mu$     & Chaotic\textsuperscript{{a,e}} (with colored noise\textsuperscript{c}),                                           \\
              & nonlinear and possibly chaotic\textsuperscript{d}                                                                 \\
    $\nu$     & Chaotic,\textsuperscript{{c}} nonstochastic\textsuperscript{a} (chaotic + Poisson noise)                          \\
    $\rho$    & Nonlinear deterministic\textsuperscript{{b,c}} (limit cycle),                                                     \\
              & nonlinear and possibly chaotic,\textsuperscript{d}                                                                \\
              & nonstochastic\textsuperscript{a} (chaotic + Poisson noise)                                                        \\
    $\phi$    & Stochastic\textsuperscript{{a,b}} (white noise\textsuperscript{c})                                                \\
    $\chi$    & Stochastic\textsuperscript{{a,b}} (white noise\textsuperscript{{c,e}})                                            \\
  \enddata
  \tablerefs{\raggedright \textsuperscript{a}\cite{misra04}, \textsuperscript{b}\cite{misra06}, \textsuperscript{c}\cite{harikrishnan11}, \textsuperscript{d}\cite{sukova16}, \textsuperscript{e}\cite{jacob16}}
  \tablecomments{All prior studies report either chaos or nonlinearity in most classes.  In contrast, our results do not indicate the presence of chaos in \emph{any} class, and the absence of stationarity in the light curves makes it difficult to convincingly assert the presence of nonlinearity through surrogate analysis (cf. Table~\ref{tab:results-summary}).}
\end{deluxetable}


\section{Comparisons with past studies}

Nonlinear time series analysis was first used to investigate the irregularity of the GRS~1915+105 light curves by \cite{misra04}.
Using an atypical implementation \citep{harikrishnan06} of the Grassberger--Procaccia algorithm, the light curve variability was classified into three distinct classes: chaotic (classes $\beta$, $\kappa$, $\lambda$, and $\mu$), stochastic (classes $\gamma$, $\phi$, and $\chi$), and ``nonstochastic'' (classes $\alpha$, $\delta$, $\theta$, $\nu$, and $\rho$).
Classes exhibiting ``some deviation'' from stochastic behavior was called nonstochastic, and it was posited that this was due to Poisson noise contamination in the light curves that were otherwise deterministic.
 \cite{misra06} complemented their earlier results with surrogate analysis and a qualitative investigation using principal component analysis.
Class $\rho$, which was earlier classified as nonstochastic, was suggested to be the manifestation of a nonlinear limit cycle, and results from surrogate analysis prompted \cite{misra06} to reclassify class $\delta$ as stochastic.
In a follow-up paper, \cite{harikrishnan11} supplemented prior studies with analysis using correlation entropy and generalized dimensions.
In this study, all classes except $\gamma$, $\delta$, $\phi$, and $\chi$ were suggested to be deterministic, with colored noise contamination in classes $\kappa$, $\lambda$, and $\mu$.
A summary of the results from these and other studies is presented in Table~\ref{tab:previous-results}.

As we have remarked, the earlier studies used an alternate algorithm by \cite{harikrishnan06} to compute the correlation dimension and the correlation entropy.
In the modified algorithm, interpoint distances (used in Equation~\ref{eq:corrsum}) are computed between a small set of randomly chosen reference points and a considerably larger set of neighbors so that most distances fall in the intermediate region where one would expect good scaling in the correlation sum plots.\footnote{In the conventional algorithm, distances between all the $\frac{1}{2}M(M - 1)$ point pairs are used.}
An estimate of $D_2$ is obtained from these distances using an automated scheme without manually observing a scaling region in the correlation sum curves.
Though the smaller set of reference points are picked randomly, this is not an unbiased estimator of $D_2$ since the larger set of neighbors are chosen only to improve scaling, and not to avoid the inherent temporal correlations along a light curve.
Thus, without explicitly employing a correction against temporal correlations, this method suffers from the same problems as the conventional method.
The inability of this method to distinguish between temporal and spatial correlations is perhaps the reason why \cite{harikrishnan06} observed a saturating $D_2$ for $1/f^{n}$-like noise, which has long-range temporal correlations (Appendix~\ref{app:dimension-entropy-estimates}).
As we have seen in the previous section, $D_2(d)$ saturates with $d$ for certain variability classes of GRS~1915+105 solely due to temporal correlations between points in their light curves.
Though the $D_2$ thus obtained could be interpreted as the fractal dimension of the embedded trajectory in the phase space, it does not represent the dimension of any underlying attractor.
Hence, in our judgment, the saturation of $D_2(d)$ reported by \cite{misra04,misra06} is a result of incorrectly including temporally correlated points while computing the correlation sum.
We also believe that failing to correct temporal correlations is again the reason \cite{harikrishnan11} found low-dimensional behavior in certain classes using correlation entropies and generalized dimensions, which also suffer from the same problem as we have shown in this paper.
It may also be mentioned that automated estimators of dimensions and entropies that do not involve manual observation of a clear scaling region, have attracted considerable criticism \citep{kantz04}.

\cite{misra04} classified cases where the computed $D_2(d)$ diverged with embedding dimension $d$, but was less than $d$, as ``nonstochastic.''
They speculated that this behavior was due to Poisson noise contamination in the light curves that were otherwise low-dimensional.
A similar argument was put forward by \cite{harikrishnan11} who suggested that contamination due to colored noise was responsible for deviation from deterministic behavior.
However, a mere deviation of the computed $D_2(d)$ from $d$ is not evidence for any nontrivial structure, as $D_2(d)$ estimated from a measured stochastic time series is in general not equal to $d$ unless a sufficiently large Theiler window $W$ is used \citep{theiler86,theiler88}.
We have also observed that $D_2(d)$ estimates obtained after using a nonzero $W$ are much higher than otherwise, and even after noise reduction, there is no significant qualitative difference in the results.
Furthermore, the issue of Poisson noise can be subtle.
As we have discussed previously, a light curve should be treated as a nonhomogeneous Poisson process in order to render any sense to the exercise of finding low-dimensional determinism in it.
Treating a light curve as a homogeneous Poisson process, as apparently done by \cite{misra04}, and by \cite{karak10} in a similar analysis of the X-ray source Cyg X-3, amounts to trivializing the problem and is debatable.

Even though the results of \cite{misra06} and \cite{harikrishnan11} do not offer convincing evidence for low-dimensional chaos, their results could be interpreted as evidence for nonlinearity in the light curves.
In both these papers, the authors reported significant differences between the dimensions and entropies of the actual light curves, and that of their surrogates representing null hypothesis NH2 of a stationary linear stochastic process.
Similar differences have also been seen in our results.
However, as we have remarked earlier, the strong lack of stationarity in the GRS~1915+105 light curves could also be the reason for these differences.
\cite{harikrishnan11} also suggested that classes $\gamma$, $\delta$, $\phi$, and $\chi$, are white noise (i.e., uncorrelated noise), since $D_2(d)$ was close to the embedding dimension $d$ for light curves from these classes.
Such a conclusion without any further check is debatable as weakly correlated time series can also have a $D_2(d)$ close to $d$.
In contrast, we have rejected null hypothesis NH1 representing uncorrelated random numbers for both $\phi$ and $\chi$ at a significance level of 0.01.
Moreover, both \cite{misra06} and \cite{polyakov12} reported nonflat power spectra for these classes, confirming that the corresponding light curves are not without significant correlations.
We have also observed considerable nonlinear structures in classes $\gamma$ and $\delta$.
The difference in the results for these classes, in our assessment, is due to the automated algorithm used to estimate $D_2$, which seems to have a bias towards dimensions obtained at very small distances.

Including temporally correlated points introduces problems in most methods of nonlinear time series analysis.
In particular, it affects recurrence analysis using both recurrence plots \citep{marwan11} and recurrence networks \citep{donner11}, where temporally correlated points are called `sojourn points' \citep{gao99}.
Without suppressing the effect of the sojourn points (for example, by using a nonzero Theiler window), they can both lead to pitfalls.
It is not apparent whether the more recent studies by \cite{sukova16} and \cite{jacob16} that report low-dimensionality in the GRS~1915+105 light curves using recurrence analysis have taken such precautions.
\cite{sukova16} reported evidence of possible nonlinear behavior and chaos in classes $\beta$, $\delta$, $\theta$, $\kappa$, $\lambda$, $\mu$, and $\rho$ after computing the correlation entropy $K_2$ from recurrence plots and comparing the results with surrogates.
However, in their study, $K_2$ was calculated only for a single value of the embedding dimension $d$ and an explicit attempt to see saturation of $K_2(d)$ with $d$ was not made.
Without establishing the invariance of the computed $K_2(d)$ with $d$, these results can only be perceived as evidence of nonlinearity in the light curves.
Furthermore, \cite{sukova16} chose the required embedding dimension for the recurrence analysis on the basis of only the first false neighbor test.
Since the embedding dimension reported by the first false neighbor test for all the GRS~1915+105 light curves is equal to the dimension that uncorrelated noise of similar lengths would have, this too cannot be taken as proof for determinism.
\cite{jacob16} examined light curves from classes $\theta$ and $\chi$ using recurrence networks and found evidence of low-dimensional chaos in class $\theta$, whereas $\chi$ was purported to be white noise.
Though we have not made an attempt to reproduce these results, in our opinion, the evidence of low-dimensional chaos in this study is because serial correlations in the data were not avoided.
As we have mentioned previously, we are also skeptical about the claim that class $\chi$ is white noise since it has nonnegligible linear correlations and a nonflat power spectrum, with a $1/f^{n}$ spectral exponent $n \approx 0.5$ \citep{polyakov12}.

Finally, it is worthwhile to single out the case of class $\rho$, also known as the `heartbeat' of GRS~1915+105 \citep{neilsen11}, which has received more attention than any other class.
The periodic bursting behavior of the $\rho$ class has been suggested to be the result of a limit cycle caused by instabilities occurring in the accretion disk \citep{taam97}.
Motivated by qualitative visualizations using principal component analysis, \cite{misra06} and \cite{harikrishnan11} also posited that the pseudoperiodic variability of class $\rho$ is due to the presence of a nonlinear limit cycle.
Though the light curves of the $\rho$ class studied in the present paper do not strictly follow deterministic dynamics, as seen from surrogate analysis (i.e., using null hypothesis NH3), a low-dimensional model of a nonlinear oscillator possibly corrupted by noise can serve as a useful approximation for its variability.
In this vein, \cite{massaro14} proposed an interesting way to model class $\rho$ variability by treating the photon flux and the accretion disk temperature as limit cycle solutions to a modified nonlinear FitzHugh--Nagumo oscillator---a commonly used biological model for neuronal activity.
Further investigations are required to make deeper connections between such low-dimensional models and the results of our study.


\section{Conclusion}

We have failed to find good evidence for low-dimensional chaos in the X-ray light curves of the microquasar GRS~1915+105 using data from \emph{RXTE} observations.
That is not to say that GRS~1915+105 does not or cannot possess states with chaotic variability.
Rather, we have simply failed to find conclusive signatures of chaos in these light curves using nonlinear time series analysis.
As we have shown, the presence of strong temporal correlations and nonstationarity in the light curves can lead to misleading evidence of low-dimensional determinism, especially when short light curves are used.
To our knowledge, this is the first study where such caveats have been considered for the nonlinear analysis of GRS~1915+105.
Evidence of nonlinearity in the GRS~1915+105 light curves we tested is also not very convincing since several of them are not long or stationary enough to faithfully represent the underlying dynamical process.
Better success at detecting low-dimensional chaos or nonlinearity may perhaps be obtained using data sets that are sufficiently long and stationary.
Considering how such light curves are presently not available for GRS~1915+1015, attempting to use nonlinear measures for classification is probably not the best approach in practice.
In view of the pessimism in the success of such methods presented in this paper, one should perhaps make use of alternative schemes of classification, e.g., \cite{polyakov12} put forward four different classes of stochastic variability (random, power law, one-scale, and two-scale) in the GRS~1915+105 light curves using flicker-noise spectroscopy.

Lastly, even if one is able to prove the presence of determinism in a light curve, extracting a low-dimensional model out of it is not straightforward and is relatively unexplored.
Though there have been attempts \citep{gouesbet91,brown94} in writing dynamical equations from a time series, naive modeling can be further complicated by the existence of multiscale dynamics \citep{cecconi12} and the question of what temporal or spatial scale such a model ought to capture.
In the present context, since photon emission mechanisms in different energy bands are expected to be different, multiscale dynamics is quite probable.
While this is a potential direction for future research, it should be borne in mind that distinguishing multiple dynamical scales using nonlinear time series analysis is a difficult task given the unsatisfactory lengths of continuous light curves available to us.
In fact, customary practical limitations on the length of data also make it difficult to distinguish between moderate- to high-dimensional chaos and stochasticity.
In such cases it is often not clear whether a chaotic or a stochastic model (or even a combination of both) is more apt.
Also, it may very well be that the best model is a set of deterministic partial differential equations that yield measurable time series, which can be stochastic or chaotic as far as tests for determinism are concerned \citep{mannattil16}.
In summary, though the results from nonlinear data analyzing techniques can often serve as useful stepping stones in understanding the light-curve variability of an astrophysical object, they are in no way a substitute for studying its physics in detail.

Having said all this, the results presented here should not be perceived as a general aversion toward using nonlinear time series analysis to study astrophysical light curves.
Careful use of these techniques, with more pragmatic expectations, will always remain a powerful tool in analyzing the ever-increasing wealth of astrophysical data.

\acknowledgments

The authors are especially grateful for the numerous discussions with Bidya Karak and Banibrata Mukhopadhyay.
We thank the editor and the anonymous reviewer for comments that led to substantial improvements in the presentation of this paper.
The authors are also thankful to Saikat Ghosh, Anand Jha, Aditya Kelkar, Anuj Nandi, Mahendra Verma, and Harshawardhan Wanare for helpful discussions.
S.C.~acknowledges financial support through the INSPIRE faculty award (DST/INSPIRE/04/2013/000365) conferred by the Indian National Science Academy (INSA) and the Department of Science and Technology (DST), India.


\appendix


\section{Time delay}
\label{app:time-delay}

Though the embedding theorems only require that the time delay $\tau > 0$ for reconstructing a chaotic attractor, for practical purposes, it is important to choose a proper $\tau$.
If $\tau$ is small, the components of the delay vector $\mathbf{y}_i^{(d)}$ are strongly correlated (\emph{redundancy}), and the reconstructed attractor lies along the phase space diagonal.
On the other hand, when $\tau$ is large, the components will have little or no dynamical correlation (\emph{irrelevance}), and the result is a structureless collection of points spread throughout the phase space \citep{galka00}.
Since both situations can lead to erroneous results, our choice of $\tau$ should balance both redundancy and irrelevance of information between the components of $\mathbf{y}_i$.
A common choice for $\tau$ is the first zero-crossing of the autocorrelation function $\langle x_i x_{i + \tau} \rangle$ as it ensures average linear independence between $\{x_i\}$ and $\{x_{i + \tau}\}$.
A more sophisticated prescription is to choose $\tau$ to be the delay at which the mutual information between $\{x_i\}$ and $\{x_{i + \tau}\}$ has its first minimum.
Mutual information, which is a statistical quantity derived from the Shannon entropy, tells us how much information can be gained about $\{x_{i + \tau}\}$ if one knows $\{x_i\}$.
Thus, when it has a minimum, we can expect the components of the delay vector to have maximum general independence.
An advantage of mutual information over other methods of choosing $\tau$ is that it also takes into consideration nonlinear dependence between the components.
Nonetheless, there is no theoretical requirement for the mutual information to have a minimum or the autocorrelation function to have a zero-crossing \citep{kantz04}.
In such situations, a rule of thumb is to choose the $1/e$ decay time of the autocorrelation function, called the autocorrelation time, as the time delay.


\section{Stationarity test}
\label{app:stationarity-test}

The stationarity test proposed by \cite{isliker93} compares the distribution of the first half of the time series with the distribution of the entire time series using a chi-squared test.
If there is no significant difference in the two distributions, we can consider the time series to be stationary.
Here we briefly summarize the method for the convenience of the readers.

We first bin the entire time series of length $N$ using a fixed set of $Q$ bins and use the resulting distribution to estimate the distribution that would result if the first half of the time series were binned using the same bins.
Our estimation is then compared with the actual distribution of the first half using a chi-squared test:
\begin{equation}
    \chi^2 = \sum_{q = 1}^{Q} \frac{\left[n_q^\text{half} - (n_q/N) N_\text{half}\right]^2}{(n_q/N) N_\text{half}}.
\end{equation}
Here $n_q$ and $n_q^\text{half}$ are the actual number of points in the $q$th bin using the full and the first half of the time series, respectively, and $N_\text{half}$ is the total number of points in the first half of the time series.
Given the value of test statistic $\chi^2$, which is a measure of the deviation between the actual and the expected distributions, we reject the null hypothesis of stationarity if the $p$-value (computed from the chi-squared distribution) is smaller than the level of significance we aim for.
In our analysis, we have used $100$ equiprobable bins for binning the light curves and have aimed for a significance level of $0.05$.


\section{Takens's estimator}
\label{app:takens-estimator}
It often helps to turn the correlation sum $C_2(r)$ into a single estimate of the correlation dimension $D_2$.
Several dimension estimators have been proposed for this purpose (see \citealt{galka00} for an extensive review).
Though such estimators have little practical relevance in computing fractal dimensions from an experimental time series, they are often used to quantify nonlinearity in the time series through surrogate analysis.
\cite{takens85} proposed an estimator that turns dimension estimation into a maximum likelihood problem involving interpoint distances.
Apart from being widely used, Takens's estimator has the least possible error for an estimate of $D_2$ computed from a set of interpoint distances.
Takens's estimate of the correlation dimension, $D_2^\text{MLE}$, can be computed from the correlation sum $C_2(r)$ using
\begin{equation}
  D_2^{\text{MLE}}(r_{\max}) = \frac{C_2(r_{{\max}})}{\int_0^{r_{{\max}} }C_2(r)/r\, \mathrm{d}r},
\end{equation}
where $r_{{\max}}$ is an upper cutoff for the interpoint distances \citep{theiler88}.
Unlike some dimension estimators, the calculation of $D_2^\text{MLE}$ does not require a lower cutoff for the distances.
Thus, all scaling information below the upper cutoff is preserved in $D_2^\text{MLE}$.
In surrogate analysis, $D_2^\text{MLE}$ is often used as a nonlinear test statistic with $r_\text{max}$ set to one-half of the standard deviation of the scalar time series \citep{kantz04}.

\section{Nonlinear noise reduction}
\label{app:nonlinear-noise-reduction}

Classical (linear) methods of noise reduction often rely on the fact that noise typically has a broadband spectra different from the underlying signal.
In contrast to classical methods, most noise reduction schemes for chaotic time series exploit the underlying deterministic nature.
Under a hypothesis of independent additive observational noise, the time series $\{x_i\}$ that we obtain after measurement can be written as:
\begin{equation}
  x_i = \hat{x}_i + \eta_i
\end{equation}
where $\hat{x}_i$ is the uncorrupted `clean' signal, and $\eta_i$ is the noise component, which we assume to be independent of the state of the system generating the time series.

In our analysis, we employ a nonlinear noise reduction scheme discussed by \cite{schreiber93}.
Aside from aspects of implementation, the algorithm works by replacing the middle coordinate of each delay vector $\mathbf{y}_i^{(d)}$ by the average of the corresponding coordinate of all the other points that are within a fixed distance $\epsilon$ from it---i.e.,
\begin{equation}
  \hat{x}_{i + m} = \frac{1}{n(U(\mathbf{y}_i^{(d)}, \epsilon))} \sum_{\mathbf{y}_j^{(d)} \in U(\mathbf{y}_i^{(d)}, \epsilon)} x_{j + m}.
\end{equation}
Here $U(\mathbf{y}_i^{(d)}, \epsilon)$ is the $\epsilon$-neighborhood around point $\mathbf{y}_i^{(d)}$ with $n(U(\mathbf{y}_i^{(d)}, \epsilon))$ points, and $m$ is the middle coordinate of $\mathbf{y}_i^{(d)}$, which is $d\tau/2$ and $(d-1)\tau/2$ for even and odd $d$ respectively.
$\epsilon$ is usually taken to be $2$--$3$ times the amplitude of the noise $\eta_i$.
The distance at which fluctuations start appearing in the correlation sum plots can be taken as an initial guess of the noise amplitude.
However, since choosing a large $\epsilon$ has the negative consequence of introducing dubious structures in the time series, we perform some consistency tests after noise reduction.
First, we ensure that the average amplitude of the noise that is effectively removed i.e., $x_i - \hat{x}_i$, is not higher than the estimates we use for $\epsilon$.
Second, since the noise is assumed to be independent of the state of the system, we also verify that there is no significant cross-correlation between the removed noise and the cleaned time series \citep{kantz04}.


\section{Potential pitfalls}
\label{app:potential-pitfalls}

Here we briefly discuss some of the pitfalls that plague the techniques used in this paper and point out some common misconceptions that arise while using them to study astrophysical light curves.
For longer general discussions, reviews by \cite{grassberger91}, \cite{schreiber99}, and \cite{bradley15} may be consulted.


\subsection{Dimension and entropy estimates}
\label{app:dimension-entropy-estimates}

When we talk about dimension and entropy estimates in nonlinear time series analysis, we are concerned with the dimensions and the entropies of the attractor \emph{represented} by the time-delayed trajectories in the phase space, rather than those of the trajectories themselves.
A trajectory bias is seen when there are either temporal correlations or nonstationarity in the time series \citep{galka00}.
In the former case, if stationarity can be guaranteed, using a judicious value for the Theiler window $W$ can ameliorate the problem considerably.
A large $W$ would not affect the analysis as the recurrence of the trajectories would ensure that the entire attractor is always sampled well.
However, if the time series is not stationary enough to ensure proper sampling, the estimated $D_2(d)$ is found to increase with $W$, and computing the actual dimension of the underlying attractor (if at all there is one) becomes impossible.
As \cite{grassberger91} noted, the failure to correct temporal correlations and account for nonstationarity is the most common reason for misleading reports of low-dimensional chaos.

An oft-cited example of correlation dimension analysis leading to spurious claims of chaos is when it is applied to noise with a power-law spectra, i.e., $1/f^n$-like noise \citep{osborne89}.
This situation arises because $1/f^n$-like noise displays long-range temporal correlations and results in nonrecurrent phase space trajectories with a fractal dimension equal to $2/(n - 1)$ \citep{feder88,theiler91}.
Thus, if temporal correlations are not corrected while calculating the correlation sum, the estimated $D_2$ saturates to the value of the dimension of the trajectory.
If carefully used, the Grassberger--Procaccia algorithm will not result in finite values of $D_2$ or $K_2$ for such stochastic processess \citep{grassberger91,kantz04}.

In Section~\ref{sec:d2}, we discussed the consistency test suggested by \cite{eckmann92} to ensure that estimated values of $D_2$ are reasonable enough.
However, such tests are often based on the scaling properties of the correlation sum curves and do not distinguish between dimensions of the trajectories and an underlying attractor.
In fact, as we saw from the results of Section~\ref{sec:data-results-discussion}, the incorrect estimates of $D_2$ obtained without removing temporal correlations pass the requirements proposed by \cite{eckmann92}.
Thus, rather than sufficient conditions, such limits should be seen as minimum necessary requirements for an accurate estimation of $D_2$.
A rule of thumb more pertinent to this issue is the suggestion that the time series should have at least 50 `structures' (i.e., full orbits or oscillations at typical time scales) so that it is able to represent the attractor well enough \citep{isliker92}.
From Figure~\ref{fig:curves} we can see that GRS~1915+105 light curves from most classes would not meet this modest requirement.
One can even go so far as to say that nonlinear time series analysis should not even be attempted on such short and nonstationary light curves.


\subsection{False nearest neighbors}
\label{app:false-nearest-neighbors}

Like other methods in nonlinear time series analysis, false nearest neighbors is not without its pitfalls.
The presence of temporal correlations in the time series introduces a trajectory bias in the method, as points that are close in time will also be close in space.
As temporal correlations are not lost at higher dimensions, near neighbors remain close, and even stochastic time series can be incorrectly judged to be deterministic.
Such a trajectory bias can be avoided by enforcing a minimum temporal separation between points before considering them as near neighbors---akin to a Theiler window used for the correlation sum \citep{fredkin95}.

Yet another problem arises if we ignore the second test (Equation~\ref{eq:fnn2}) for short data sets.
As we have remarked in Section~\ref{sec:false-nearest-neighbors}, the fraction of false nearest neighbors becomes zero for noisy data sets if the second test is ignored.
In particular, for uncorrelated noise of length $N$, the average distance between near neighbors increases as $N^{-1/d}$ and one would always obtain a zero false neighbor fraction for $d > \log_A{N}$ using the first test.
Thus, for a given time series of length $N$, an estimated minimum embedding dimension larger than $\log_A{N}$ is meaningless without the second test \citep{hegger99}.
Since $N \leq 7200$ for light curves from all the GRS~1915+105 classes analyzed here, with $A = 10.0$, we would expect the first test to report negligible false neighbors after an embedding dimension $d > \log_{10}{7200} \approx 4$, as we saw from the actual results in Section~\ref{sec:data-results-discussion}.


\subsection{Surrogate analysis}
\label{app:surrogate-analysis}

In surrogate analysis, a successful rejection of the null hypothesis only means that it cannot account for the observed properties of the time series, and the rejection should never be taken as evidence of low-dimensional determinism.
(It should also be emphasized that \citeauthor{theiler92}~[\citeyear{theiler92}] originally introduced surrogate analysis as a test for nonlinearity, and not for determinism or chaos.)
For example, if the time series is the output of a nonlinear stochastic process (which is neither chaotic nor deterministic), we can always reject the null hypothesis of linearly correlated noise.
We should only consider a time series to be low-dimensional if it has unambiguous signatures of determinism such as a clear scaling region in the correlation sum plots with a rapidly converging value of $D_2(d)$.

One should also be cautious while interpreting the results of surrogate analysis when the original time series is nonstationary.
Constrained realization algorithms to generate surrogates assume that the input time series is stationary (as it is supposed in the null hypothesis), and the resulting surrogate data sets also tend to be stationary.
For example, such algorithms would not be able to imitate nonstationary jumps and bursts in a light curve, and the resulting surrogates would look very different from it visually.
Hence, if the original time series is nonstationary (as has been the case for several GRS~1915+105 light curves), then the null hypothesis is already violated, and a rejection does not tell us much---not even nonlinearity.
Though there are methods to include nonstationarity in the null hypothesis, their success is limited and requires one to know the exact nature of nonstationarity present in the time series \citep{schreiber00}.


\end{document}